\documentclass[%
 reprint,
superscriptaddress,
 amsmath,amssymb,
 aps,
]{revtex4-2}

\usepackage{graphicx}
\usepackage{dcolumn}
\usepackage{gensymb}
\usepackage{bm}

\begin{document}

\preprint{APS/123-QED}

\title{Phase-Transition Induced Magnetic Domain Evolution and Magnetization Dynamics in FePt/FeRh Bilayers for Advanced Heat-Assisted Magnetic Recording}

\author{Saroj K. Mishra}
\affiliation{National Institute for Materials Science, Tsukuba, Ibaraki 305-0047, Japan}
\affiliation{Nanomagnetism and Microscopy Laboratory, Department of Physics, Indian Institute of Technology Hyderabad, Sangareddy 502285, Telangana, India}

\author{Y. Sasaki}
\affiliation{National Institute for Materials Science, Tsukuba, Ibaraki 305-0047, Japan}

\author{S. Isogami}
\affiliation{National Institute for Materials Science, Tsukuba, Ibaraki 305-0047, Japan}

\author{I. Suzuki}
\affiliation{Western Digital Technologies GK, Fujisawa, Japan.}

\author{Keerthana P}
\affiliation{Nanomagnetism and Microscopy Laboratory, Department of Physics, Indian Institute of Technology Hyderabad, Sangareddy 502285, Telangana, India}

\author{J. Mohanty}
\affiliation{Nanomagnetism and Microscopy Laboratory, Department of Physics, Indian Institute of Technology Hyderabad, Sangareddy 502285, Telangana, India}

\author{Y. K. Takahashi}
\email{TAKAHASHI.Yukiko@nims.go.jp}
\affiliation{National Institute for Materials Science, Tsukuba, Ibaraki 305-0047, Japan}

\date{\today}

\begin{abstract}
Achieving ultrahigh recording densities with low power consumption is a central challenge for next-generation heat-assisted magnetic recording (HAMR), as conventional $L1_0$-FePt media require intense laser heating due to their high coercivity ($\mu_0H_c$) and high Curie temperature ($\sim$700 K). Here, we address this issue using FePt/FeRh bilayers, where the antiferromagnetic-to-ferromagnetic transition of FeRh near 350 K generates strong interfacial exchange coupling that assists magnetization switching in the FePt layer. Magnetometry measurements reveal a 40\% reduction in $\mu_0H_c$ from 300 K to 400 K in the bilayer, compared to only 8\% in single-layer FePt. Temperature-dependent MFM directly captures phase-transition-induced domain evolution, showing a $30\%$ reduction in domain size and enhanced phase contrast. TR-MOKE measurements reveal only a minor ($\sim$0.4 T) modification of the effective anisotropy field during phase-transition, confirming that the intrinsic anisotropy of FePt remains largely preserved. These results demonstrate that the reduction in $\mu_0H_c$ in FePt/FeRh bilayers is primarily governed by phase-transition-induced domain-wall mobility coupled with interfacial magnetic interactions, rather than by intrinsic anisotropy softening. This mechanism provides a pathway toward efficient magnetization switching under reduced thermal load, making FePt/FeRh heterostructures promising candidates for advanced HAMR media.
\end{abstract}

\maketitle


\section{\label{sec:level1}INTRODUCTION}
The rapid expansion of digital technologies and data-centric applications has driven a significant increase in global demand for information storage \cite{ozdemir2018birth,rydning2018digitization}. Large-scale data centers, which form the backbone of modern cloud computing and artificial intelligence infrastructures, rely predominantly on hard disk drives (HDDs) owing to their high storage capacity, nonvolatility, long-term reliability, and low cost per bit \cite{zha2025data,yang1999comprehensive}. Sustained increases in areal density are therefore essential to meet future storage requirements while simultaneously limiting data center energy footprints. Current commercial HDDs operate at areal densities of approximately 1.5 $Tbit/in^2$, and next-generation technologies aim to exceed 4 $Tbit/in^2$ and beyond \cite{albuquerque2021hdd,wu2013hamr,marchon2013head,parker2014hamr}. Achieving such ultrahigh recording densities necessitates a substantial reduction in the size of the magnetic grains that collectively store each bit of information. However, decreasing the grain volume inevitably compromises thermal stability, which is governed by the formula $K_uV/K_BT > 60$, where  $K_u$ is the uniaxial magnetic anisotropy, $V$ is the grain volume,  $K_B$ is the Boltzmann constant, and $T$ is the temperature \cite{weller2000high}. Conventional CoCrPt-based perpendicular magnetic recording media cannot satisfy this condition at grain sizes below approximately 5 nm \cite{qin2009development}. $L1_0$-ordered FePt has therefore emerged as a key material for next-generation recording media due to its exceptionally large $K_u$, which ensures long-term thermal stability even at nanometer-scale grain sizes \cite{chen2008review}. Despite this advantage, the very high $K_u$ of $L1_0$-FePt results in a large coercive field ($\mu_0H_c$), making magnetization switching difficult with conventional write heads.
In addition, reducing the grain size to increase areal density decreases the magnetic volume, lowering the signal-to-noise ratio (SNR) due to reduced readback signal amplitude and increased statistical fluctuations. This fundamental trade-off among thermal stability, writeability, and SNR is commonly known as the magnetic recording trilemma \cite{wang2015overcoming}. 

Heat-assisted magnetic recording (HAMR) has been developed as a practical solution to this challenge by temporarily heating the recording media to a temperature close to FePt's Curie temperature ($T_c$) during the write process \cite{weller2016fept}. Localized laser heating reduces the $K_u$ and $\mu_0H_c$ of FePt, enabling magnetization reversal with a moderate write field, followed by rapid cooling to restore thermal stability, ensuring long-term data retention. Although HAMR has reached commercial deployment, further improvements in areal density and device reliability remain limited by the high operating temperatures required for writing \cite{wang2013hamr}. In FePt-based media, the Curie temperature approaches 700 K \cite{yk2022microstructure}, and repeated exposure to such extreme thermal conditions can induce material degradation, structural deformation, and damage to both the recording media and the write head components \cite{trinh2020temperature,ma2018study,xu2012thermal}. In addition, the lubricants that protect the recording medium's surface can deteriorate when subjected to repeated or prolonged thermal cycling \cite{zhang2006lubrication}. Consequently, reducing the required thermal load during HAMR writing is a critical materials and device challenge for next-generation HDD technologies. In this context, alternative approaches that lower the switching field without relying solely on near $T_c$ heating are highly desirable, such as thermal spin torque HAMR systems, in which spin torques driven by self-formed temperature gradients effectively reduce the switching field \cite{isogami2025thermal}. The other approach to address this challenge is to exploit interfacial magnetic interactions in engineered heterostructures, where additional degrees of freedom can assist magnetization reversal without sacrificing long-term thermal stability. 

\begin{figure}[htbp]
\centering
\includegraphics[width = 3.3in ,height=1.9in]{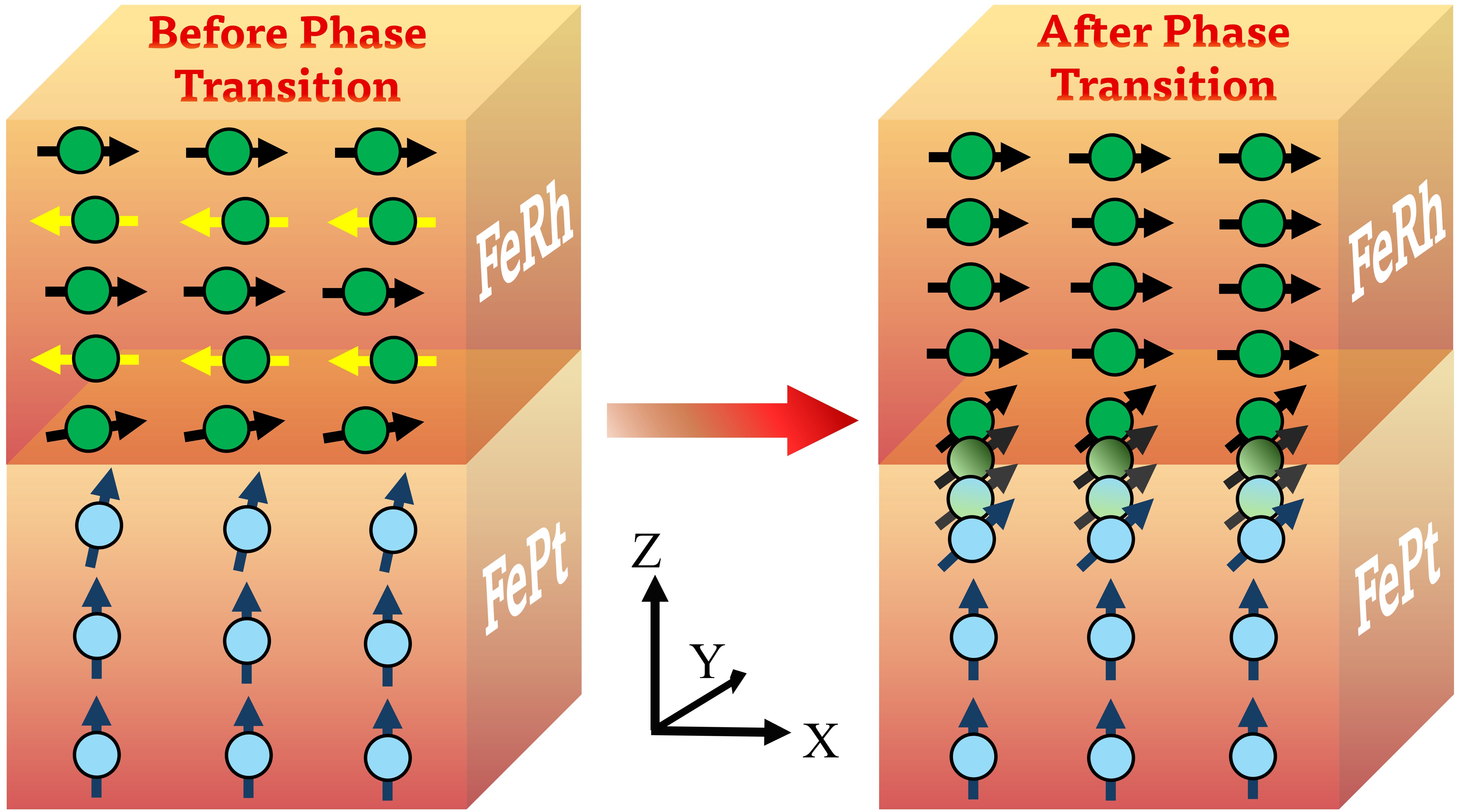}
\caption{Schematic illustration of Thermally driven exchange-coupling-assisted magnetization switching in FePt/FeRh bilayer. Before the phase transition, the Néel vector in the FeRh layer points along in-plane direction, resulting in weak interfacial exchange coupling with FePt. Upon heating above the phase transition temperature, FeRh transforms to the FM phase, establishing strong interfacial exchange coupling that facilitates magnetization switching in the FePt layer.}
\end{figure}

In this work, we investigate FePt/FeRh bilayers with perpendicular magnetic anisotropy (PMA), as schematically illustrated in Fig. 1, to elucidate the mechanism of thermally driven magnetization behavior. FeRh is known to undergo a first-order magnetic phase transition from an antiferromagnetic (AFM) ground state at room temperature (RT) to a ferromagnetic (FM) state upon heating to approximately 350 K \cite{chirkova2017effect,suzuki2009stability}. At RT, the AFM FeRh layer carries no net magnetization, thereby preserving the high thermal stability of the FePt recording layer. However, when the temperature exceeds the phase-transition threshold, FeRh develops an FM moment and becomes strongly magnetic-coupled to FePt at the interface. This thermally induced evolution of interfacial magnetic coupling modifies the magnetization reversal process in FePt, enabling switching at lower temperatures compared to conventional HAMR, where reversal primarily relies on the reduction of $K_u$
near the $T_c$ of FePt.

Early studies by $Thiele~et~al.$ demonstrated that exchange-coupled FePt/FeRh bilayers exhibit a pronounced reduction in $\mu_0H_c$ upon heating, arising from the thermally induced AFM-to-FM phase transition of FeRh and the formation of an exchange-spring system \cite{thiele2003ferh}. Subsequent investigations by $Lu~et~al.$ provided detailed insight into the temperature-dependent magnetization reversal processes in epitaxial FePt/FeRh bilayers, revealing a transition from exchange bias to exchange-spring behavior and a two-step reversal mechanism governed by domain wall (DW) nucleation and propagation across the soft-hard magnetic interface \cite{lu2012magnetic}. Complementary work by $Nguyen~et~al.$ focused on exchange-bias effects in FePt/FeRh bilayers, reporting a strong $K_u$ and a thickness-dependent modulation of $\mu_0H_c$, and emphasizing the importance of interfacial coupling in determining the magnetic response \cite{nam2009exchange}. $Koyama~et~al$. further clarified the roles of FeRh thickness and nucleation fields, showing that a significant reduction in $\mu_0H_c$ occurs only when the FeRh layer effectively participates in magnetization twisting during its magnetic phase transition \cite{koyama2005reduction}. While these studies collectively established the fundamental temperature-dependent magnetic behavior and the reduction in $\mu_0H_c$ in FePt/FeRh systems, which primarily focused on quasi-static magnetic properties and did not explicitly address thermally driven exchange coupling as a dynamic switching mechanism under HAMR-relevant conditions. 

In this work, we extend beyond $\mu_0H_c$ reduction and conventional exchange-spring phenomenology by directly demonstrating the origin of enhanced $\mu_0H_c$ reduction in FePt/FeRh heterostructures, supported by a systematic investigation of their structural, magnetic, microscopic, and dynamic properties. Temperature-dependent magnetometry and nanoscale magnetic imaging reveal that the thermally driven phase transition of FeRh significantly reduces $\mu_0H_c$ by promoting domain-wall mobility and inducing domain evolution. Time-resolved pump-probe measurements show only a minor variation in the effective anisotropy field of FePt, confirming that its intrinsic magnetic stability remains largely preserved. These findings demonstrate that the observed reduction in $\mu_0H_c$ is governed by phase-transition-induced domain dynamics and interfacial magnetic interactions, enabling efficient magnetization switching under reduced thermal load and providing a viable pathway toward next-generation low-power HAMR media.

\section{EXPERIMENTAL DETAILS}
\subsection{Sample fabrication}
Single-crystal MgO(001) substrates were cleaned with acetone and ethanol, followed by in situ thermal flashing at 700 $\celsius$ in the magnetron sputtering chamber to remove residual surface contaminants such as adsorbed hydrocarbons, moisture, and surface oxides. The base pressure of the deposition chamber was maintained at approximately $1 \times 10^{-6}$ Pa. A 10 nm thick FePt single-layer film was first deposited using a stoichiometric $Fe_{50}Pt_{50}$ alloy target at a substrate temperature of 700 $\celsius$  to promote $L1_0$ chemical ordering and PMA. Subsequently, a 20 nm thick FeRh layer was deposited on top of the FePt layer by co-sputtering at the same substrate temperature. The composition of the FeRh layer ($Fe_{49}Rh_{51}$) was confirmed by electron probe microanalysis (EPMA). The crystalline structure and degree of $L1_0$ chemical ordering were characterized by X-ray diffraction (XRD) using a Rigaku SmartLab diffractometer with a Cu $K_{\alpha}$ radiation source. Magnetic properties were measured using a superconducting quantum interference device (SQUID) magnetometer (Quantum Design).

\subsection{Microstructure analysis}
Microstructural and compositional analyses were performed using scanning transmission electron microscopy (STEM) combined with energy-dispersive X-ray spectroscopy (EDS) in a Spectra Ultra STEM (Thermo Fisher Scientific) equipped with a probe aberration corrector. Cross-sectional electron-transparent lamellae were prepared by a focused ion beam (FIB) lift-out technique using an FEI Helios Nanolab 650 system. To minimize ion-beam-induced damage during FIB milling, a protective gold (Au) layer was deposited on the film surface prior to specimen preparation.

\subsection{Magnetic domain observation}
Temperature-dependent magnetic domain imaging was carried out using a magnetic force microscope (MFM, Park NX10) equipped with a temperature-controlled stage. The heating stage enables precise temperature regulation up to 250 $\celsius$ with a resolution of 0.1 $\celsius$. All measurements were conducted on a vibration-isolated platform to ensure mechanical stability. To minimize probe–sample magnetic interactions and avoid tip-induced modifications of the domain structure, a low-moment magnetic probe (Bruker MESP-LM-V2) was employed, and a constant lift height of 40 nm was maintained during magnetic imaging.

\subsection{Magnetization dynamics}
Temperature-dependent magnetization dynamics were investigated by all-optical time-resolved magneto-optical Kerr effect (TRMOKE) using a femtosecond laser system with a pulse duration of 230 fs, a center wavelength of 1028 nm, and a repetition rate of 10 kHz. The wavelength of a probe laser pulse was converted to 528 nm using a $\beta-BaB_2O_4$ (BBO) crystal, and the resulting beam was irradiated onto the sample surface. Its Kerr rotation angle was detected by a balanced photodetector with a quarter-wave plate and a Wollaston prism. The amplitude of the pump laser pulse was modulated with 570 Hz using a mechanical chopper, and the probe pulse polarization change was detected using a Lock-in amplifier. The time delay between the probe and pump pulses was controlled by a mechanical delay line. During the TRMOKE measurement, sample temperature was controlled by a homemade thermal heater inside the superconducting magnet chamber, and a magnetic field up to $\mu_0 H$ = 7.0 T was applied with a field angle of $80^{\degree}$ with respect to the film normal (Further details are described in a previous report \cite{sasaki2023thermal}).

\section{RESULTS AND DISCUSSION}
Fig. 2 presents the out-of-plane (OOP) XRD patterns of the FePt single layer and the FePt/FeRh bilayer. In both films, the presence of the $L1_0$-ordered FePt phase is clearly confirmed by the observation of the (001) superlattice reflection along with the (002) fundamental peak, indicating a high degree of chemical ordering and a strong (001) texture. For the bilayer, additional diffraction peaks corresponding to the B2-ordered FeRh layer are observed, where the (001) superlattice and (002) fundamental reflections evidence epitaxial growth of FeRh on the underlying FePt layer.

\begin{figure}[h]
\centering
\includegraphics[width = 3.2in ,height=3.6in]{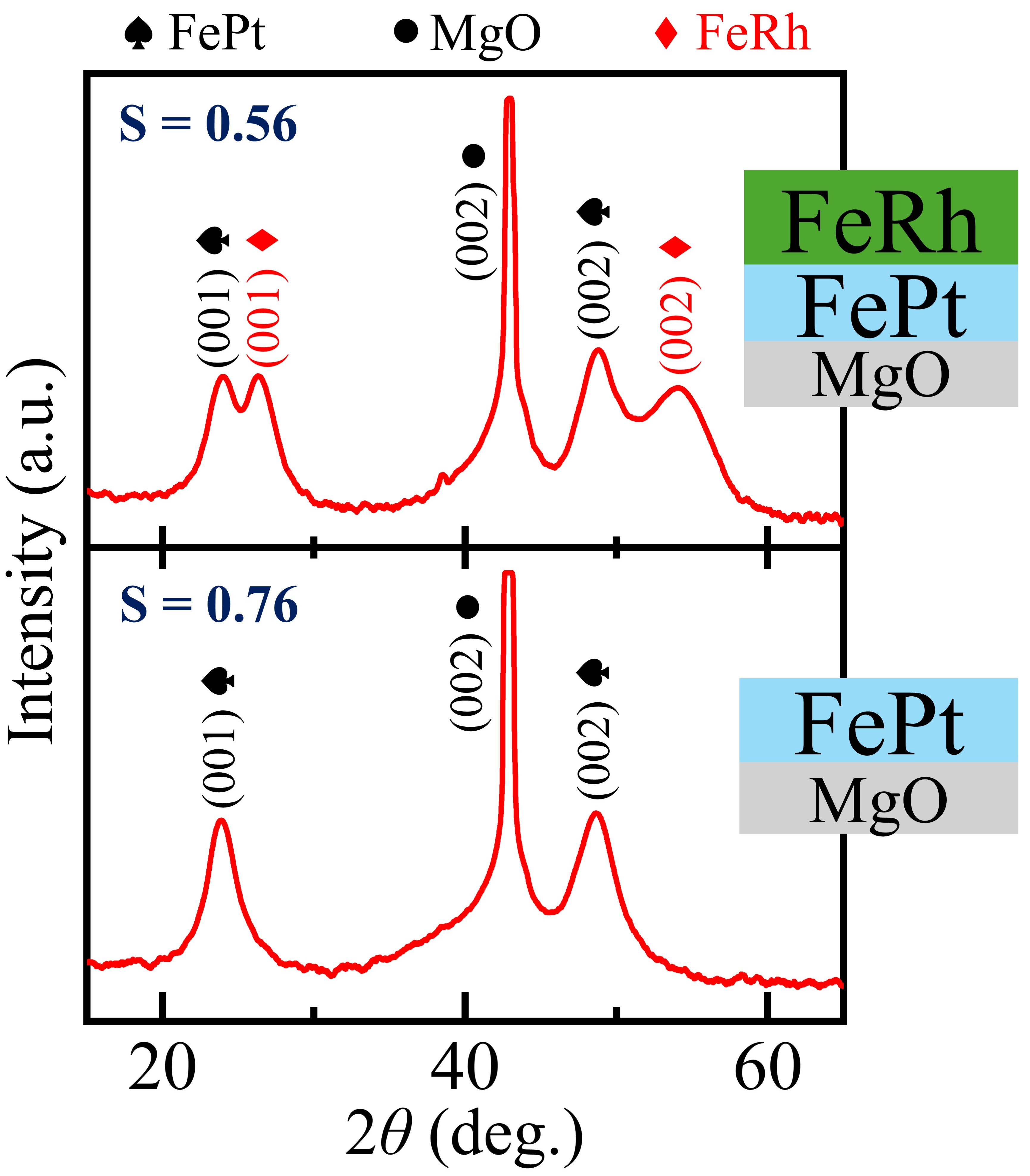}
\caption{X-ray diffraction (XRD) patterns of the FePt single layer and the FePt/FeRh bilayer, confirming the $L1_0$ chemical ordering of FePt.}
\end{figure}

\begin{figure*}[htbp]
\centering
\includegraphics[width = 6.9in ,height=3.9in]{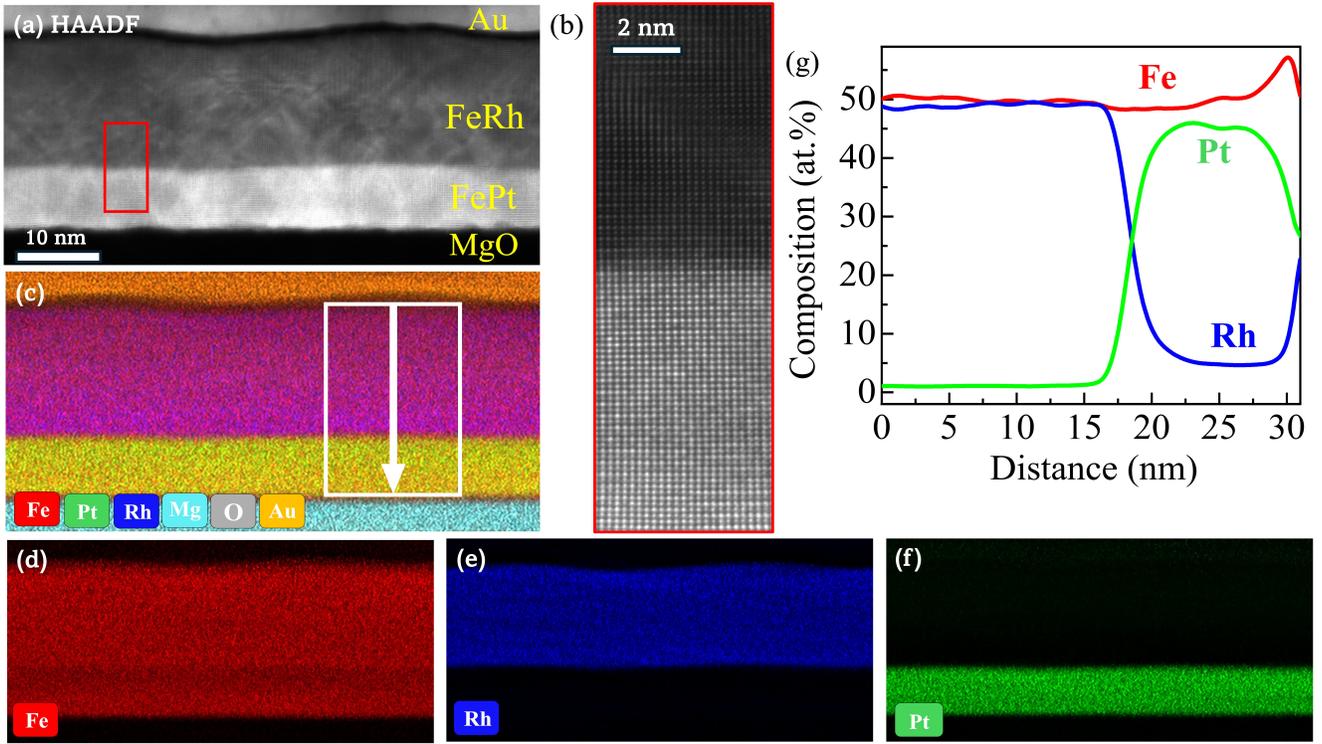}
\caption{Depth-resolved microstructural characterization of the FePt/FeRh bilayer. (a) Cross-sectional HAADF-STEM image revealing the well-defined layered structure. (b) Composite EDS elemental map showing the depth distribution of Fe, Pt, Rh, Mg, and O across the film stack. (c-e) Corresponding individual EDS maps of Fe, Rh, and Pt, demonstrating compositional uniformity and sharp interfaces between the layers. (f) High-resolution TEM image acquired from the region highlighted by the red box in (a), highlighting the crystalline quality of the bilayer. (g) Compositional line profiles extracted along the depth (z-axis) direction.}
\end{figure*}

When comparing the XRD results of the FePt single layer and the FePt/FeRh bilayer, the ordering parameter S of the FePt layer decreases from 0.76 in the single layer to 0.56 in the bilayer. Similarly, the $B2$ ordering parameter of FeRh reduces from 0.78 in the single layer (Fig. S1) to 0.61 in the bilayer, indicating that the formation of the bilayer affects the chemical ordering of both layers. This reduction can be primarily attributed to interfacial intermixing during high-temperature growth, which introduces local compositional inhomogeneity and suppresses long-range ordering near the interface \cite{chen2013effects}. In addition, temperature-dependent XRD measurements of the FeRh single layer (Fig. S2) reveal lattice expansion during heating, leading to partial strain relaxation, which may further influence the structural ordering in the bilayer system. Similar phenomena have been reported in FePt films where the introduction of intermediate or underlayers disrupts chemical ordering kinetics and grain growth, resulting in lower order parameters compared to isolated FePt films \cite{negi2023formation,ogawa2024exchange}. 

To examine the interfacial structure and possible interdiffusion in the FePt/FeRh heterostructure, cross-sectional high-angle annular dark-field (HAADF) STEM imaging was performed, as shown in Fig. 3(a). Owing to the strong atomic-number (Z) contrast, the FePt layer appears brighter than the FeRh layer, reflecting the higher average atomic number of Pt compared to Rh. The image clearly resolves the layered structure, allowing distinction between the FePt and FeRh regions and their interface. High-resolution STEM imaging of the region marked by the red box in Fig. 3(a) is presented in Fig. 3(b), enabling a more detailed view of the lattice arrangement across the interface. The lattice-resolved image confirms epitaxial growth of both layers along the OOP (001) direction. To investigate interfacial diffusion, EDS compositional analysis was performed. The composite elemental map across the full film stack is shown in Fig. 3(c), while the individual elemental distributions of Fe, Rh, and Pt are displayed in Figs. 3(d)-3(f), respectively. These maps reveal a well-defined layer contrast across the film. However, since 2D EDS maps collect signals along the electron-beam direction, they cannot clearly distinguish subtle compositional variations along the film normal (depth) direction. Therefore, laterally averaged compositional line profiles were extracted along the growth direction (z-axis), as shown in Fig. 3(g). The profiles reveal a discernible diffusion of Fe and Rh into the FePt layer. Such interdiffusion is sufficient to locally disrupt the FePt chemical ordering, thereby providing a microscopic origin for the reduced $L1_0$ order parameter. 

Fig. 4(a,b) presents the magnetic hysteresis loops measured along the OOP and IP directions for the FePt single layer and the FePt/FeRh bilayer. The FePt single layer exhibits a square OOP loop with a $\mu_0H_c$ of $\sim$0.15 T, a $K_u$ of 2.4 MJ/m$^3$, and negligible IP remanence, confirming strong PMA originating from the well-ordered $L1_0$ phase. Upon introducing the FeRh overlayer, the bilayer retains PMA with a similar $\mu_0H_c$, but shows a reduced $K_u$ of 1.6 MJ/m$^3$ along with an increased IP remanence at RT. These changes can be attributed to interfacial magnetic and structural effects introduced by the FeRh layer. As revealed by XRD and STEM-EDS analysis, the reduction in $L1_0$ chemical ordering and interfacial interdiffusion can lead to a decrease in the $K_u$, which might allow partial canting of spins away from the OOP direction. In addition, exchange interactions at the AFM/FM interface, mediated by the possible presence of uncompensated FeRh spins, can induce non-collinear spin configurations near the interface \cite{mitsumata2006uncompensated,liu2008configuration}. These effects collectively introduce a finite IP magnetization component, resulting in the observed enhancement of IP remanence in the bilayer.

Figs. 4(c,d) present the temperature-dependent magnetization (M-T) curves measured with the magnetic field applied perpendicular to the film plane for the FeRh single layer and the FePt/FeRh bilayer. A clear AFM-to-FM phase transition associated with the FeRh layer is observed in both samples. From here, we will define the transition temperature, $T_{tr}$, as the temperature at which the magnetization reaches half of its total change during heating. For the single FeRh layer, the transition is relatively sharp, with $T_{tr}$ = 330 K, whereas in the bilayer it becomes broader and shifts to approximately 350 K. This behavior can be attributed to interfacial coupling and structural constraints imposed by the adjacent FePt layer. In the bilayer, the FeRh layer experiences an additional interfacial exchange coupling with the FM FePt, which alters the free-energy landscape of the AFM-FM transition. This coupling can be represented as an effective exchange field ($H_{\mathrm{ex}}$) that favors FM alignment near the interface. Accordingly, the total free energy includes an interfacial exchange term ($E_{\mathrm{int}} \sim -J_{\mathrm{int}} \mathbf{M}{\mathrm{FePt}} \cdot \mathbf{M}{\mathrm{FeRh}}$), which alters the thermodynamic balance between AFM and FM states and shifts the $T_{tr}$. As a result, a higher thermal energy is required to complete the transition compared to uncoupled FeRh.  Previous studies have shown that exchange interactions at phase boundaries and interfaces significantly influence the evolution of the AFM–FM transition and stabilize FM regions near interfaces \cite{massey2020phase}. Furthermore, the thermal hysteresis width of the transition, defined as the temperature difference between the heating and cooling branches, is reduced from about 26 K in the single FeRh layer to approximately 15 K in the FePt/FeRh bilayer. Such an effect can arise from superheating and supercooling of competing magnetic phases \cite{lu2012magnetic}. The reduced hysteresis width in the bilayer suggests that exchange coupling to the FePt layer lowers the nucleation barrier for the FM phase in FeRh, facilitating a more reversible and energetically favorable transition.

\begin{figure}[htbp]
\centering
\includegraphics[width = 3.3in ,height=4.2in]{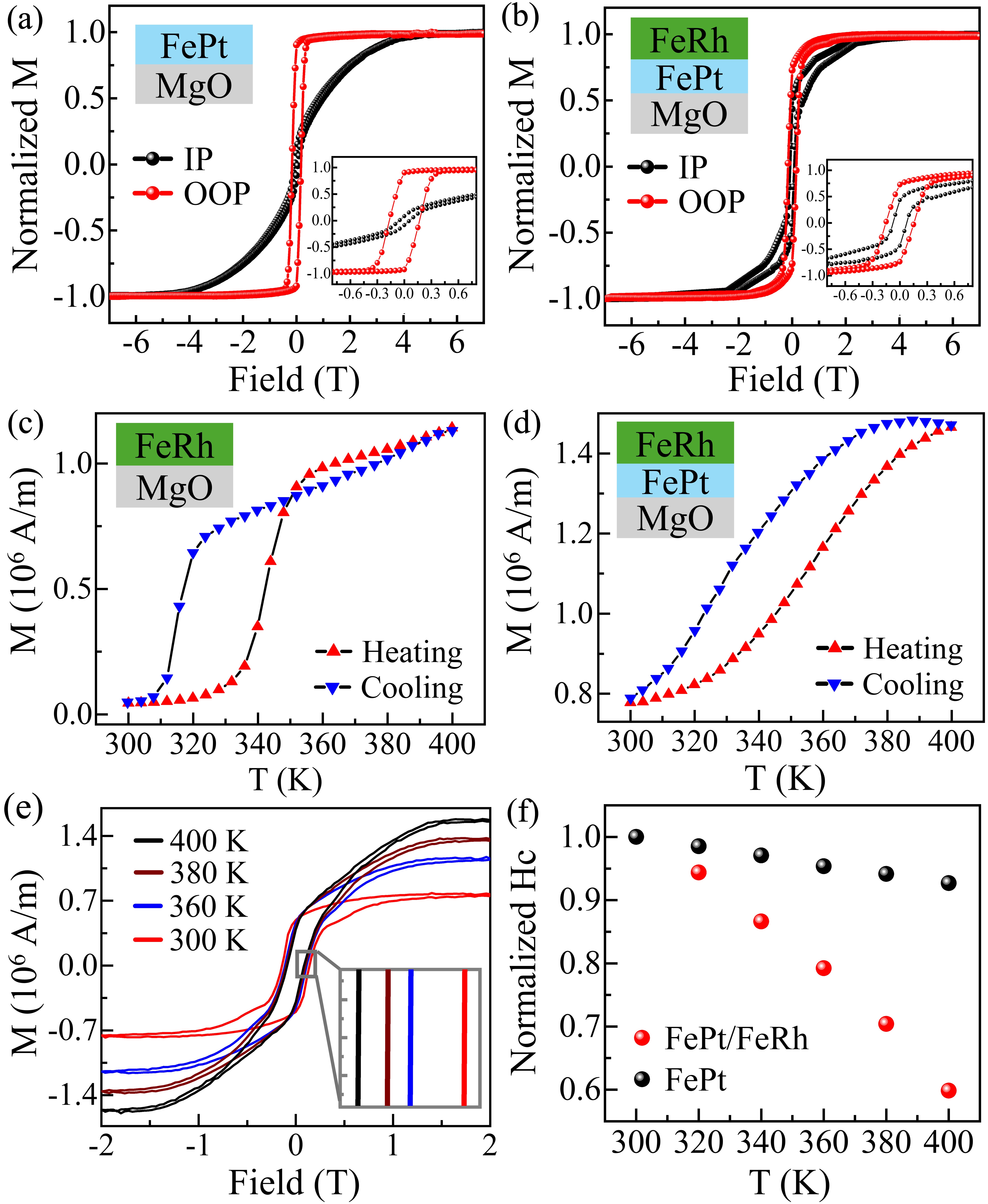}
\caption{ In-plane and out-of-plane magnetic hysteresis loops of the (a) FePt single layer and (b) FePt/FeRh bilayer films. Temperature-dependent magnetization measurements illustrating the thermally driven magnetic response of the (c) FePt single layer and (d) FePt/FeRh bilayer. (e) Temperature-dependent OOP magnetization curves showing a thermal enhancement of saturation magnetization accompanied by a gradual reduction in coercivity; the inset (gray box) highlights the evolution of coercivity. (f) Direct comparison of the temperature-induced coercivity reduction in the FePt single layer and the FePt/FeRh bilayer.}
\end{figure}

The temperature-dependent OOP hysteresis loops of the FePt/FeRh bilayer are presented in Fig. 4(e). As the temperature increases from 300 to 400 K, the saturation magnetization increases from approximately $0.8 \times 10^6 A/m$ to $1.6 \times 10^6 A/m$, reflecting the progressive AFM-FM phase transition of the FeRh layer. The corresponding evolution of the $\mu_0H_c$ is highlighted in the inset. For comparison, Fig. 4(f) summarizes the temperature dependence of $\mu_0H_c$ for both the FePt single layer and the FePt/FeRh bilayer. The temperature-dependent hysteresis loops for the FePt and FeRh single layers are shown in Figs. S3 and S4 of the supporting information. While the FePt single layer exhibits only a modest $\mu_0H_c$ reduction of approximately $\sim$8\% (from 0.15 T to 0.138 T) over the same temperature range, the FePt/FeRh bilayer shows a pronounced decrease in $\mu_0H_c$ from 0.15 T to 0.09 T, corresponding to a reduction of about $\sim$40\%. This substantial $\mu_0H_c$ reduction in the bilayer is primarily attributed to the thermally induced AFM-FM phase transition of FeRh, which leads to the emergence of a FM moment and strengthens the interfacial exchange coupling with the hard FePt layer.

\begin{figure*}[ht]
\centering
\includegraphics[width = 6.1in ,height=6.1in]{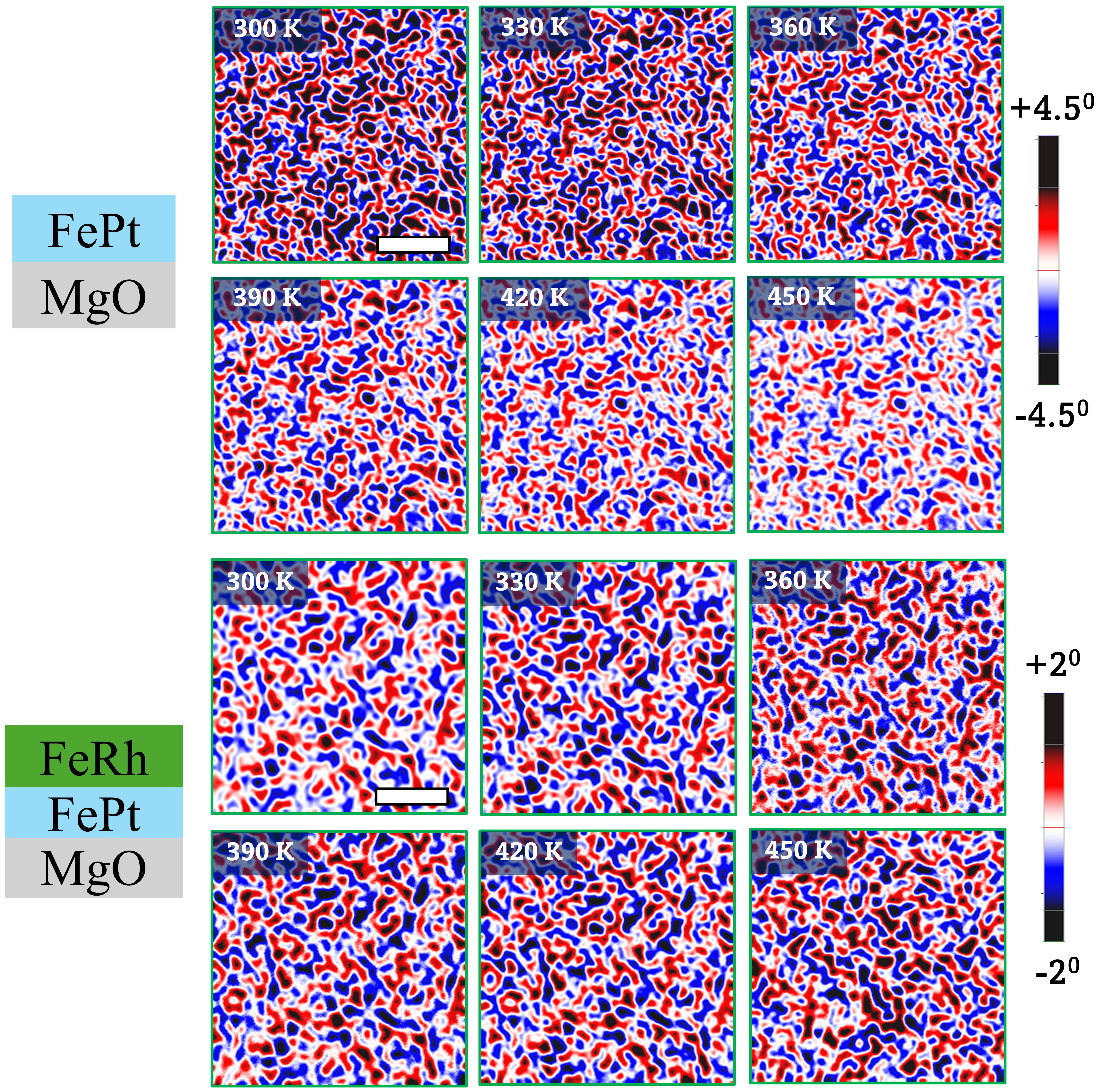}
\caption{Normalized MFM images of the FePt single layer and FePt/FeRh bilayer acquired at various temperatures. The corresponding temperature is indicated in the upper-left corner of each panel. All images share a uniform scale bar of 1 $\mu m$.}
\end{figure*}

To obtain a clearer physical picture of the $\mu_0H_c$ reduction, we performed temperature-dependent MFM measurements to track the evolution of magnetic domains. Prior to the MFM measurements, the film was first demagnetized to achieve a near-equilibrium magnetic state that minimizes the total magnetic energy. Fig. 5 presents representative MFM images of the FePt single layer and the FePt/FeRh bilayer acquired at various temperatures. The corresponding topographic images are provided in Fig. S5. The phase signal in the MFM images, represented by the color scale, reflects the magnetic force component associated with the OOP component of the sample magnetization ($F_{mag} \propto M^2$) \cite{martin1987magnetic}. To enable a reliable comparison across different temperatures, the phase scale was kept constant across all images, set to $\pm 4.5^{\degree}$ for the FePt single layer and $\pm 2^{\degree}$ for the FePt/FeRh bilayer. In the color maps, bright red and blue regions represent magnetic domains with magnetization oriented along the upward and downward OOP directions, respectively. The white contrast observed at the boundaries between these regions corresponds to DWs, where the OOP magnetization component approaches zero, resulting in a minimized stray-field gradient.

\begin{figure}[htbp]
\centering
\includegraphics[width = 2.5in,height=3.9in]{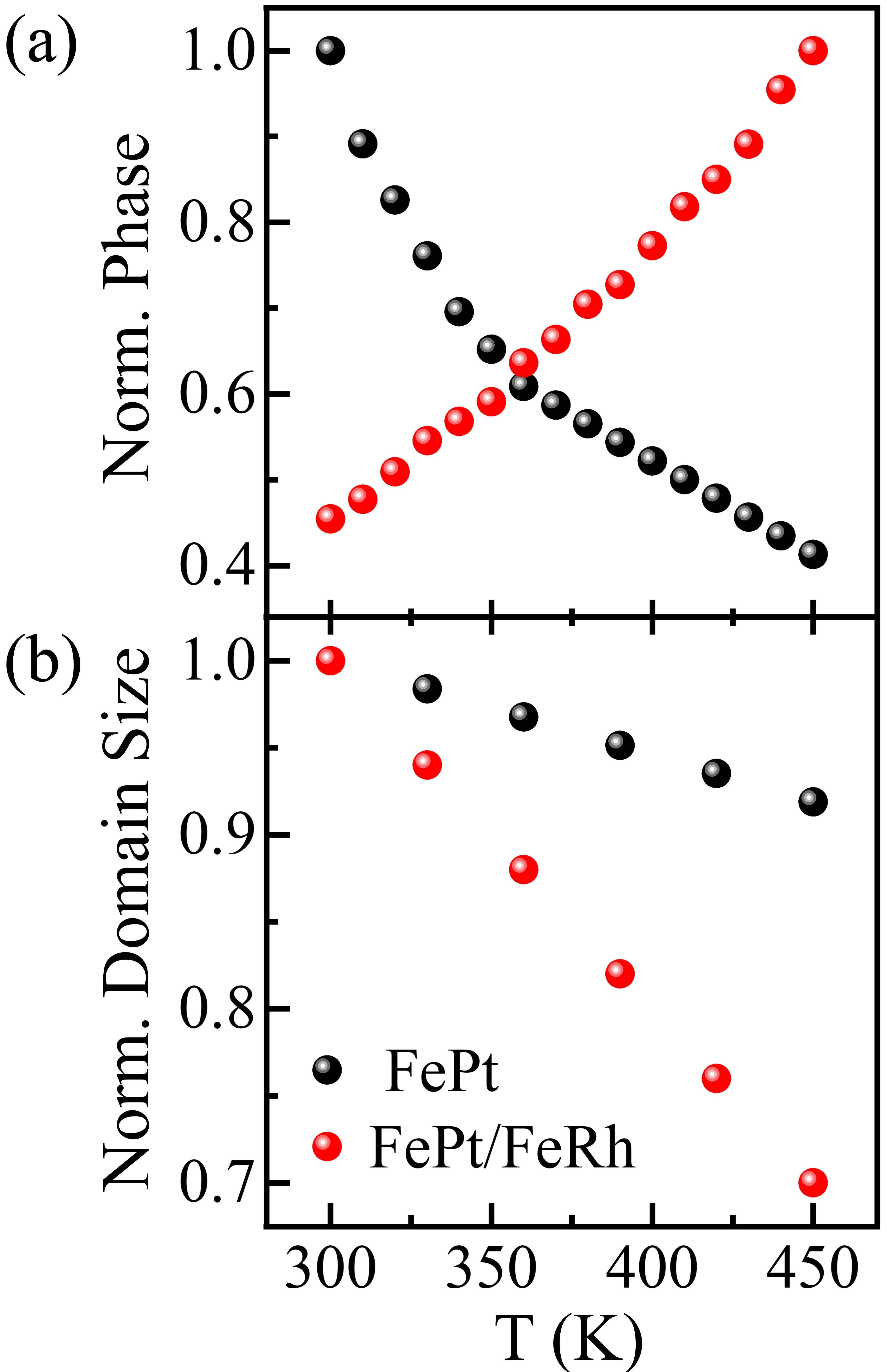}
\caption{Temperature dependence of the normalized (a) phase contrast and (b) average domain size for the FePt single layer and FePt/FeRh bilayer, extracted from MFM measurements}
\end{figure}

At 300 K, the FePt single layer exhibits a well-defined maze-like domain structure, characteristic of perpendicular magnetic films where exchange, anisotropy, and magnetostatic energies compete to minimize the total free energy. The strong PMA stabilizes narrow domains with well-defined boundaries. As the temperature increases, a gradual reduction in magnetic phase contrast is observed [Fig. 6(a)], accompanied by an apparent broadening of the white DW regions. At 450 K, the disappearance of distinct up- and down-oriented domains and the dominance of white contrast indicate thermally induced reductions in spontaneous magnetization, leading to decreased DW energy and enhanced DW mobility, which results in diffuse and less well-defined domain structures \cite{almeida2017quantitative}. In contrast, the FePt/FeRh bilayer exhibits a lower domain density and lower phase contrast ($\pm 2^{\degree}$) with relatively broader white-contrast regions at 300 K. This behavior can be correlated with the reduced $K_u$ and the emergence of a finite IP magnetization component inferred from the magnetization curve [Fig. 4(b)]. The reduced $K_u$ leads to an increase in DW width, since the wall width scales as $\delta = \pi \sqrt{A/K_u}$ \cite{thiaville2012dynamics,kumar2024engineering}, resulting in more pronounced white contrast in MFM images, which is associated with IP magnetization within the DWs. Upon heating to 450 K, a progressive increase in phase contrast [Fig. 6(a)] and domain density is observed. This behavior directly correlates with the AFM-to-FM phase transition of FeRh. As the temperature increases, FM regions nucleate in FeRh, thereby enhancing interfacial exchange coupling with FePt and increasing the system's magnetization and $K_u$ [Fig. 4(e)]. This reduces DW width (white contrast) and sharpens domain boundaries, while thermally activated DW mobility facilitates domain nucleation and reconfiguration.

To quantitatively compare the temperature-induced evolution of domain morphology in the FePt single layer and the FePt/FeRh bilayer, the normalized average domain size was plotted as a function of temperature, as shown in Fig. 6(b). The FePt single layer shows only a modest reduction in domain size ($\sim$8\%, from 185 nm to 170 nm) with increasing temperature, consistent with the gradual decrease in magnetostatic energy. However, the FePt/FeRh bilayer exhibits a much stronger domain refinement ($\sim$30\%, from 250 nm to 175 nm). The characteristic domain size was extracted from the power spectral density (PSD) analysis (Fig. S6), where the Lorentzian-fitted peak position corresponds to the inverse of the domain periodicity, and half of this periodicity defines the average domain width \cite{mishra2026growth}. 

The pronounced decrease in domain size in the bilayer originates from a temperature-driven rebalancing of magnetic energy contributions. As FeRh undergoes its AFM-to-FM transition, the emergence of a FM moment increases the magnetostatic energy and $K_u$ of the coupled system. To reduce the associated stray-field energy, the system stabilizes a higher domain density, leading to a smaller equilibrium domain size. Simultaneously, thermal relaxation of strain in the FeRh layer (Fig. S2), together with the reduction in $\mu_0H_c$ [Fig. 4(e)], can reduce defect-mediated DW pinning. This reduction in pinning enhances DW mobility, enabling the domain configuration to reorganize toward a lower-energy state with finer domain structures. These observations collectively provide direct microscopic evidence of thermally activated interfacial magnetic coupling in the FePt/FeRh bilayer, which facilitates magnetization reversal through domain restructuring.

To quantify the modulation of the effective PMA field $\mu_0 H_\mathrm{k}^{\mathrm{eff}}$ in the FePt layer and separate it from the large $\mu_0H_c$ reduction in the bilayer during the phase transition, we employed all-optical TR-MOKE measurements. In this configuration, the probe beam was incident from the substrate side, making the measurement more sensitive to the FePt layer despite the presence of the FePt/FeRh heterostructure. Typical wave forms at ambient temperature for various magnetic fields ($\mu_0 H$) are shown in Fig. 7(a). After pump-pulse irradiation of the film surface at an optical delay time of $\Delta t$ = 0 ps, ultrafast demagnetization, indicated by a rapid decrease in the MOKE signal $\Delta \phi_k$, was observed. During the magnetization recovery process, a damped oscillatory signal corresponding to the magnetization precessional motion was observed. As indicated by the red solid curves, these oscillational signals were analyzed by fitting them using the least squares method through the following equation: 
\begin{equation}
\begin{split}
    \Delta \phi_k = A_1 . \mathrm{exp}(- \frac{\Delta t}{\tau}).\mathrm{Sin}(2\pi f \Delta t + \lambda) + B_0 + \\ B_1.\mathrm{exp}(- \frac{\Delta t}{{\tau}'}),
    \end{split}
\end{equation}
where the background signal was described using a signal offset $B_0$, signal amplitude of magnetization recovery process $B_1$, and its decay time $\tau'$. First term describes the oscillational signal, where an amplitude of precession motion $A_1$, decay time of the precession $\tau$, precession frequency $f$, and phase $\lambda$. Phase and frequency varied with magnetic fields. The Fast-Fourier-Transforms of TRMOKE data at $\mu_0 H$ = 4.0 T with different temperatures $T$ were shown in Fig. 7(b), and its peak frequency slightly decreased with increasing $T$ owing to the reduction of the effective magnetic field in the FePt layer. 

\begin{figure}[htbp]
\centering
\includegraphics[width = 3in ,height=3.7in]{TRMOKE.jpg}
\caption{ (a) Typical wave forms of TRMOKE data at temperature of T= 300 K with various magnetic fields $\mu_0H$ at magnetic field angle of $80^{\degree}$ with respect to the film normal. Red solid curves show fitting results. (b) Fast-Fourier-Transform spectrum of TRMOKE data with various temperatures T at $\mu_0 H$ = 4.0 T.}
\end{figure}

\begin{figure}[htbp]
\centering
\includegraphics[width = 3in ,height=5.3in]{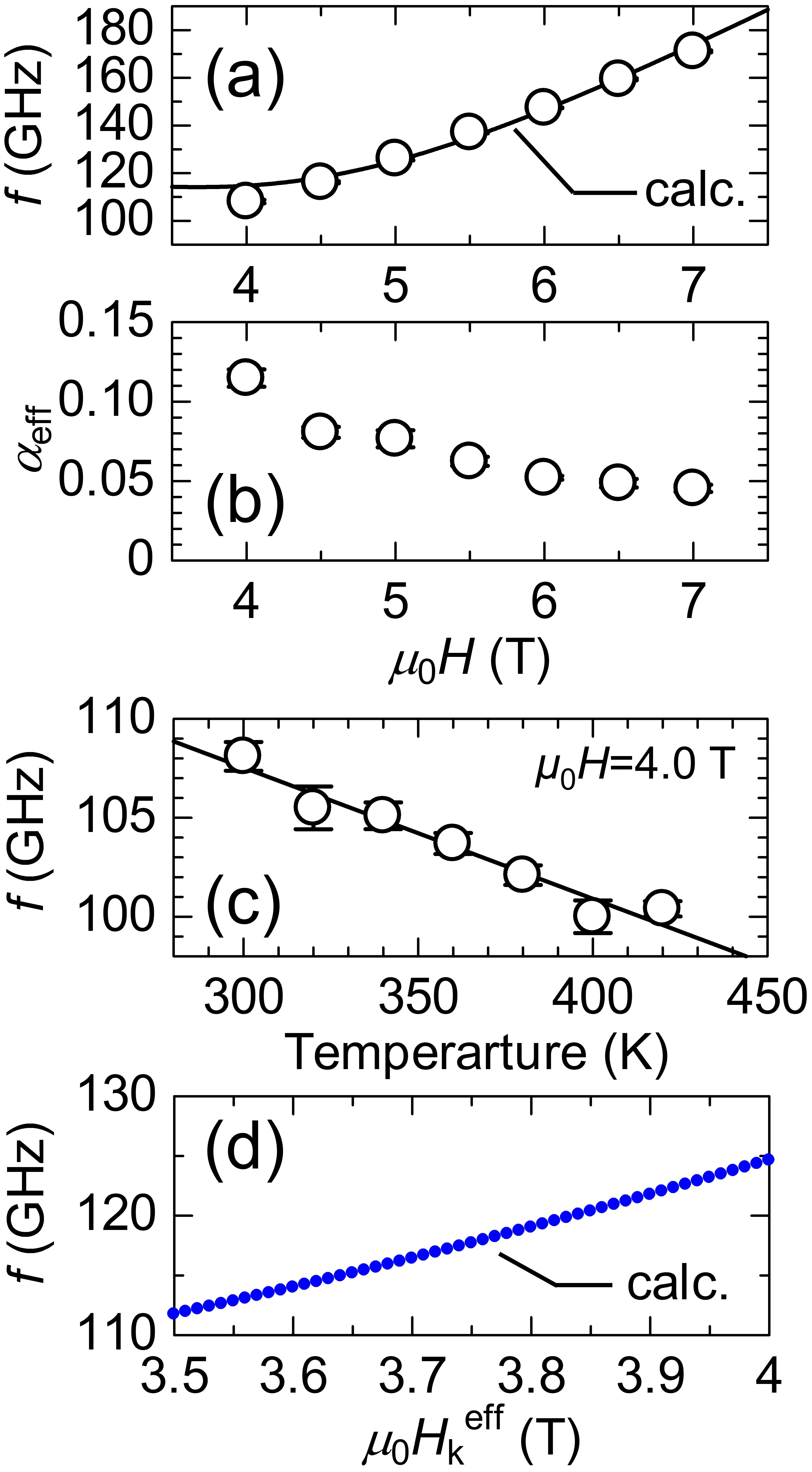}
\caption{ (a) Precession frequency f and (b) effective magnetic damping constant $\alpha_\mathrm{{eff}}$ as a function of magnetic field $\mu_0 H$ evaluated from TRMOKE data at temperature of T = 300 K. The solid curve shows the theoretical fit. (c) f as a function of temperature for the TRMOKE data with $\mu_0H$ = 4.0 T. A Solid curve shows a linear fit. (d) Theoretical calculation result of f as a function of effective perpendicular magnetic anisotropy field $\mu_0 H_\mathrm{k}^\mathrm{{eff}}$ at $\mu_0 H$ = 4.0 T.}
\end{figure}

The effective PMA field $\mu_0 H_\mathrm{k}^\mathrm{{eff}}$ can be evaluated from the magnetic field dependence of $f$ at ambient temperature as shown in Fig. 8(a). The $f$ versus $\mu_0H$ data are fitted (indicated by solid curve) using Kittel’s formula by taking into account the first-order uniaxial PMA as described by the following equations \cite{sasaki2022magnetization,iihama2014gilbert}:
\begin{equation}
f = \frac{\mu_0 \gamma}{2\pi}\sqrt{H_1H_2}
\end{equation}
\begin{equation}
\mu_0H_1 = \mu_0H ~ \mathrm{cos}(\theta_H-\theta_M) + \mu_0 H_\mathrm{k}^\mathrm{{eff}}~\mathrm{cos}^2(\theta_M)
\end{equation}
\begin{equation}
\mu_0H_2 = \mu_0H ~ \mathrm{cos}(\theta_H-\theta_M) + \mu_0 H_\mathrm{k}^\mathrm{{eff}}~\mathrm{cos}(2\theta_M)
\end{equation}
where $\gamma$ is defined as $\gamma$  = $g\mu_B/\hbar$ with Lande’s g-factor g, Bohr magneton $\mu_B$, and Dirac constant $\hbar$. $\theta_M$ is the equilibrium angle of magnetization with respect to the film normal and was determined from the following relationship between the PMA and Zeeman energy.
\begin{equation}
2\mu_0H~\mathrm{sin}(\theta_H-\theta_M) + \mu_0 H_\mathrm{k}^\mathrm{{eff}}~\mathrm{sin}(2\theta_M) = 0
\end{equation}
The fitting yields a g-value of 2.4 and $\mu_0 H_\mathrm{k}^\mathrm{{eff}}$ = 3.8 T, respectively. Fig. 8(b) shows the external magnetic field dependence of the effective magnetic damping constant $\alpha_\mathrm{{eff}}$, which was evaluated from the precession frequency and lifetime using the following equation:
\begin{equation}
\alpha_{eff} = \frac{1}{2\pi f\tau}.
\end{equation}
$\alpha_\mathrm{{eff}}$ shows a minimum value of = 0.045 at $\mu_0 H$ = 7.0 T, and this value is consistent with previous standard $L1_0$-FePt films deposited on MgO substrates \cite{bentley2025development,ma2015role}. This relatively large $\mu_0 H_\mathrm{k}^\mathrm{{eff}}$ indicates that our observation has been performed only in the FePt layer, although the stack contains FePt/FeRh heterostructure, because of the signal sensitivity at the surface scaled by optical penetration depth. 

Fig. 8(c) shows the temperature dependence of $f$ at $\mu_0 H$ = 4.0 T. The Solid line is the linear fit using the least-squares method, with a slope of -0.066 GHz/K. Theoretical calculation of $f$ with g-value of 2.4 and $\mu_0 H_\mathrm{k}^\mathrm{{eff}}$ = 3.8 T represents the frequency shift of 25 GHz/T as shown in Fig. 8(d). Therefore, we conclude that $\mu_0 H_\mathrm{k}^{\mathrm{eff}}$ is modulated by only $\sim$0.4 T ($\sim$10\% of 3.8 T), as estimated from the frequency shift, whereas the $\mu_0H_c$ reduction is much larger ($\sim$40\%) [Fig. 4(f)] over the temperature range of 300 K to 400 K. This clear difference indicates that the pronounced reduction in $\mu_0H_c$ observed in the FePt/FeRh bilayer cannot be explained by changes in the intrinsic magnetic properties of the FePt layer. The  frequency shift evaluation procedure is described in detail in a previous report \cite{sasaki2017all}. Since $\mu_0 H_\mathrm{k}^\mathrm{{eff}}$ incorporates contributions from the intrinsic PMA, exchange field, and demagnetization field, the limited modulation indicates that the PMA of the FePt layer remains largely stable with temperature. Therefore, the pronounced thermal reduction of the $\mu_0H_c$ in the FePt/FeRh film cannot be attributed solely to intrinsic anisotropy softening. Instead, it likely originates from interlayer effects, such as thermally activated exchange coupling with the FM FeRh phase, dipolar interactions, or interfacial spin reorientation during the AFM-FM transition of FeRh. This decoupling between stable PMA and strongly reduced $\mu_0H_c$ is particularly advantageous for HAMR, enabling a substantial reduction in the switching field while preserving the high anisotropy required for thermal stability.

\section{CONCLUSION}
In this work, we systematically investigated the structural, magnetic, microscopic, and dynamic properties of FePt/FeRh bilayers to elucidate the origin of the temperature-dependent reduction in $\mu_0H_c$. XRD was employed to confirm the epitaxial growth and chemical ordering of the $L1_0$-FePt and B2-FeRh phases, while cross-sectional TEM provided insight into the crystalline quality and interfacial structure, revealing slight interdiffusion that reduced $L1_0$ ordering. Magnetic measurements established PMA at room temperature and revealed a pronounced 40\% reduction in $\mu_0H_c$ between 300 K and 400 K, coinciding with the AFM-FM transition of FeRh. To uncover the microscopic origin of this behavior, temperature-dependent MFM was performed, which reveals a ~30\% reduction in domain size and increased phase contrast in the bilayer, indicating dipolar-energy rebalancing and enhanced DW mobility. Complementary TR-MOKE measurements probed the magnetization dynamics and effective anisotropy, showing only a minor ($\sim$0.4 T) variation during the phase transition, thereby confirming that the intrinsic anisotropy of FePt remains largely preserved. These results demonstrate that the $\mu_0H_c$ reduction in FePt/FeRh bilayers is primarily governed by phase-transition-induced magnetic domain-wall evolution and interfacial exchange coupling, rather than intrinsic anisotropy softening. This study provides key insight into thermally driven magnetic reconfiguration in continuous FePt layers. Building on this understanding, a futuristic direction is to extend this concept to FePt granular media, which offers a promising pathway toward practical low-power HAMR applications.

\begin{acknowledgments}
This work was supported by the International Cooperative Graduate Program (ICGP). A part of this work was supported by the Electron Microscopy Unit of the National Institute for Materials Science (NIMS). This work was supported in part by the JST-CREST (JPMJC22C3), MEXT program: Data Creation and Utilization-Type Material Research and Development Project (JPMXP1122715503) and ASRC. The authors thank Ms. Yukie Mori (NIMS) for FIB sample preparation. The authors acknowledge JX Advanced Metals Corporation for providing the FePt target used in this work.
\end{acknowledgments}

\nocite{*}

\bibliography{apssamp}

\end{document}